\documentclass[%
 reprint,
 amsmath,amssymb,
 aps,
]{revtex4-2}

\usepackage{graphicx}
\usepackage{dcolumn}
\usepackage{bm}

\begin{document}

\preprint{APS/123-QED}

\title{Hopfions of massive gauge bosons in early universe}

\author{M. Bousder}
 \email{mostafa.bousder@fsr.um5.ac.ma}
 \affiliation{Laboratory of Condensend Matter and Interdisplinary Sciences, Department of physics,\\
 Faculty of Sciences, Mohammed V University in Rabat, Morocco
}
\author{H. Ez-Zahraouy}
 \email{h.ezzahraouy@um5r.ac.ma}
 \affiliation{Laboratory of Condensend Matter and Interdisplinary Sciences, Department of physics,\\
 Faculty of Sciences, Mohammed V University in Rabat, Morocco
}

\date{\today}

\begin{abstract}
This letter presents a novel model that characterizes the curvature of
space-time, influenced by a massive gauge field in the early universe. This
curvature can lead to a multitude of observations, including the Hubble
tension issue and the isotropic stochastic gravitational-wave background. We
introduce, for the first time, the concept of gauge field Hopfions, which
exist in the space-time. We further investigate how hopfions can influence Hubble parameter values. Our findings open the door to utilizing hopfions as a topological source which links both gravitation and the gauge field.
\end{abstract}

\maketitle


\textit{Introduction}. During the inflationary period, the interplay between
coupled axion and SU(2) gauge fields \cite{IN1,IN2} offers a diverse range
of phenomena, distinguishing itself from the characteristics observed in
conventional single scalar field inflation models \cite{IN3,IN4}. Notably,
this coupling gives rise to intriguing outcomes, such as the generation of a
stochastic background comprising chiral gravitational waves \cite{IN5,IN6},
characterized by their non-Gaussian nature \cite{IN7,IN8}. Moreover, recent
studies \cite{IN9,IN10,IN11} have demonstrated that the production of these
chiral gravitational waves results in a non-zero parity-violating
gravitational anomaly. This anomaly breaks lepton number symmetry and helps
to explain the observed baryon asymmetry in the universe. Massless gauge
bosons generate gravitational wave signals, as studied in \cite{IN12,IN13}.
Gravitational waves from tensor perturbations are a common outcome of simple
inflation models. In \cite{JCAP4}, they show how massive gauge bosons also
produce gravitational wave signals through tensor fluctuations during
inflation. Massive particles are usually hard to produce during inflation
due to a Boltzmann-like factor $exp(-\pi m/H)$, where $m$ is the particle's
mass and $H$ is the Hubble rate, the inflationary scale. This reduces the
signals at the cosmological collider. Within the realm of topological
solitons, magnetic skyrmions are two-dimensional entities resembling
particles, characterized by a continuous twist of magnetization. On the
other hand, magnetic Hopfions are three-dimensional structures that can be
constructed from a closed loop of intertwined skyrmion strings. Theoretical
frameworks propose that magnetic Hopfions can achieve stability in
frustrated or chiral magnetic systems, and that target skymions can be
converted into Hopfions by adjusting their perpendicular magnetic
anisotropy. However, experimental confirmation of these theories remains
elusive to date \cite{NUT1,PRL1,PHY1}.\newline
Recently, there has been a significant orientation towards the study of
magnetic Hopfions \cite{IN14,IN15,IN16,IN17,IN18}. Our aim in this letter is
to present, for the first time, the concept of gauge field Hopfions. These
exist in space-time in the early universe and could potentially resolve the
tension issue of Hubble and the signal from pulsar timing arrays.

\begin{figure}
\centering\includegraphics[width=9cm]{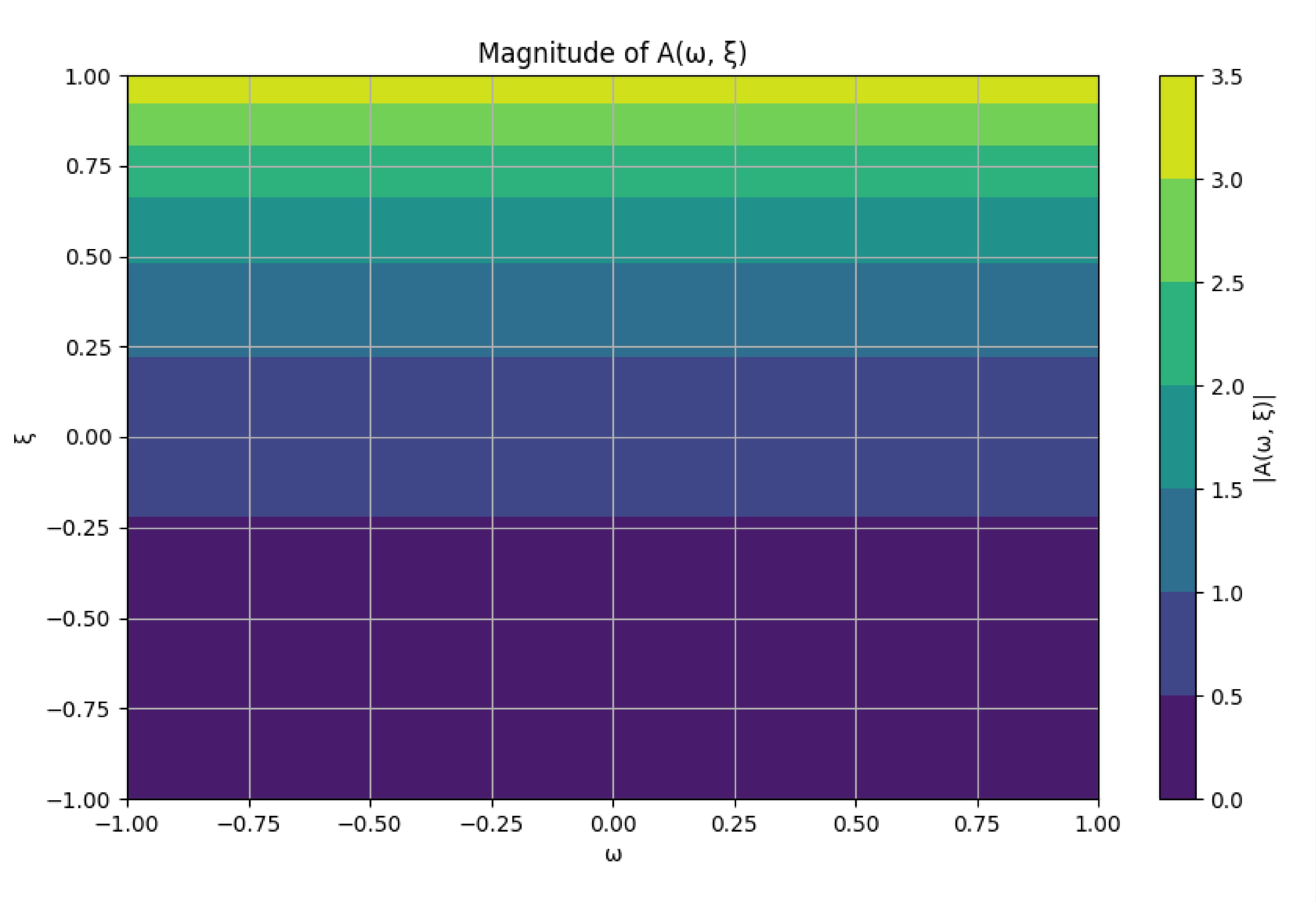}
\caption{The plot shows how the gauge field $A_{\mathbf{+}}$\ changes with
respect to $\left( \protect\omega ,\protect\xi \right) $. Parametersare
defined. as $z=k=\protect\tau =1$ and $\protect\tau _{0}=1.$}
\label{F1}
\end{figure}

\begin{figure}
\centering\includegraphics[width=9cm]{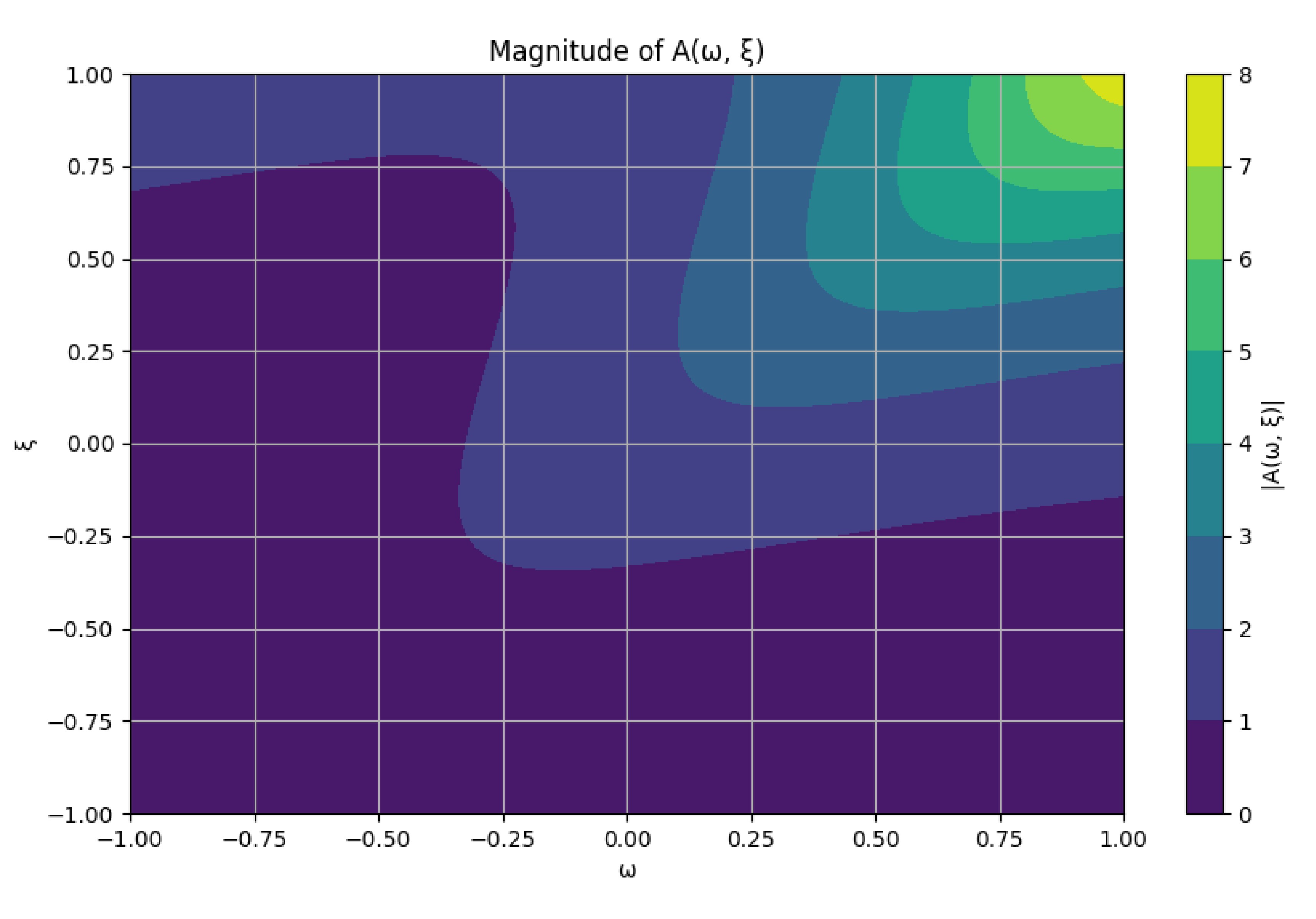}
\caption{The plot shows how the gauge field $A_{\mathbf{+}}$\ changes with
respect to $\left( \protect\omega ,\protect\xi \right) $. Parametersare
defined. as $z=k=\protect\tau =1$ and $\protect\tau _{0}=1.5$.}
\label{F2}
\end{figure}

\begin{figure}
\centering\includegraphics[width=9cm]{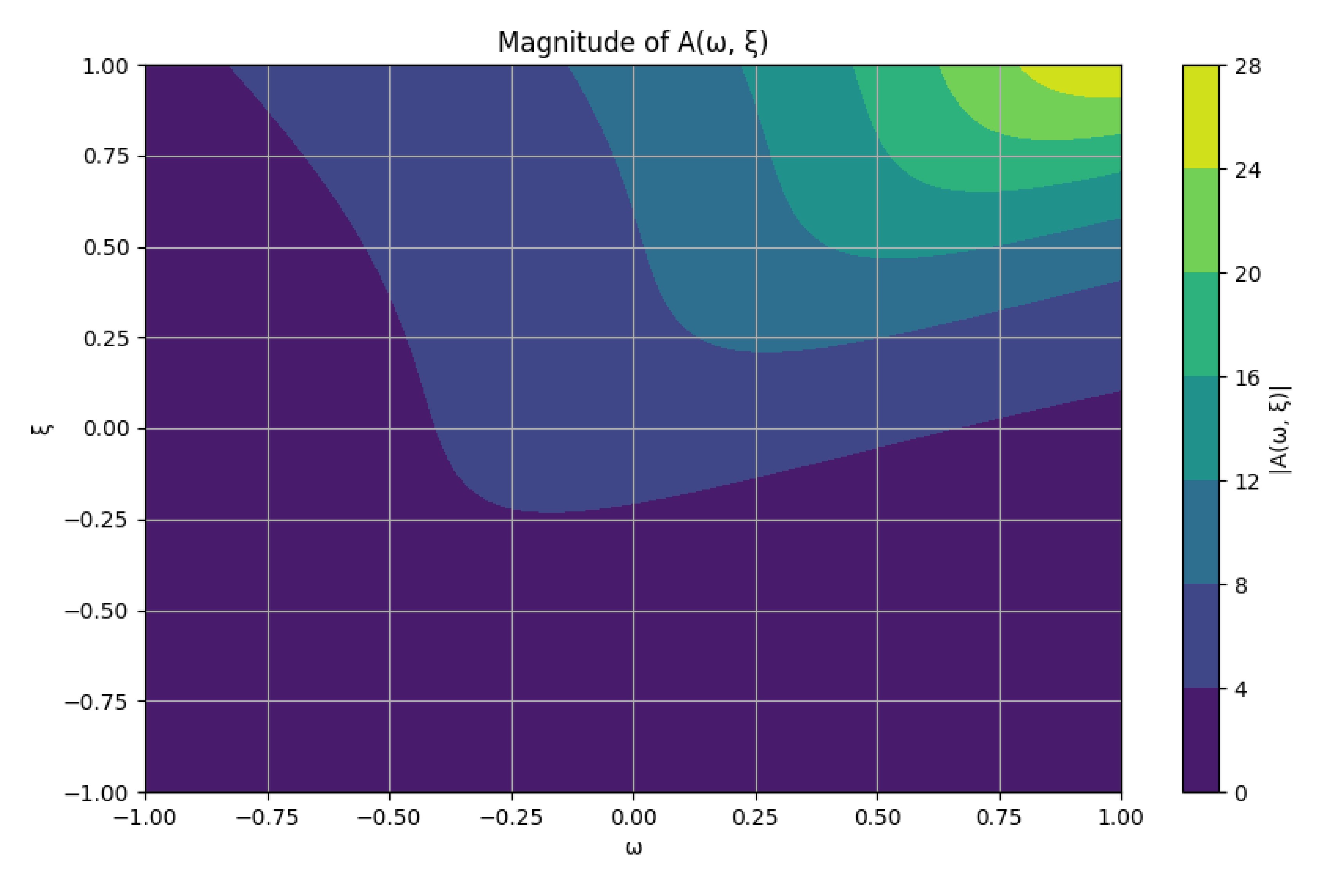}
\caption{The plot shows how the gauge field $A_{\mathbf{+}}$\ changes with
respect to $\left( \protect\omega ,\protect\xi \right) $. Parametersare
defined. as $z=k=\protect\tau =1$ and $\protect\tau _{0}=3$.}
\label{F3}
\end{figure}

\textit{Inflation with massive gauge bosons}. We study how to generate a
force field with mass using the Chern-Simons interaction. We use a U(1)
gauge field $A_{\mu }$ and add a term $\phi F_{\mu \nu }\tilde{F}^{\mu \nu }$
that connects it with the inflaton field $\phi $ \cite{PRD3,JCAP4}%
\begin{eqnarray}
S &=&\int d^{4}x\sqrt{-g}(-\frac{1}{2}\partial _{\mu }\phi \partial ^{\mu
}\phi -V\left( \phi \right) -\frac{1}{4}F_{\mu \nu }F^{\mu \nu }.  \label{m1}
\\
&&-\frac{\sigma }{4\Lambda }\phi F_{\mu \nu }\tilde{F}^{\mu \nu }+\frac{1}{2}%
m_{A}^{2}A^{\mu }A_{\mu }).  \notag
\end{eqnarray}%
Here, $\Lambda $ represents the mass scale that suppresses
higher-dimensional operators in the theory, while $\sigma $ is the
dimensionless coupling of the inflaton-gauge field. Here $A_{\mu }$ is U(1)
gauge field $A_{\mu }$ and its field strength tensor $F_{\mu \nu }=\partial
_{\mu }A_{\nu }-\partial _{\nu }A_{\mu }$. We also have its dual $\tilde{F}%
^{\mu \nu }=\frac{1}{2}\frac{\epsilon ^{\mu \nu \alpha \beta }}{\sqrt{-g}}%
F_{\alpha \beta }$. We focus on the key role of $\frac{1}{2}m_{A}^{2}A^{\mu
}A_{\mu }$, which connects the gauge field and the gravitational waves
through the mass $m_{A}$. The energy of the Chern-Simons interaction is $-%
\frac{\sigma }{8\Lambda }\int d^{3}xe^{-i\mathbf{k\cdot x}}\delta \phi
\epsilon ^{\mu \nu \alpha \beta }F_{\mu \nu }F_{\alpha \beta }$. We consider
a homogeneous and isotropic universe described by the Friedmann-Lema\^{\i}%
tre-Robertson-Walker (FLRW) metric $ds^{2}=-dt^{2}+a^{2}(t)g_{ij}dx^{i}dx^{j}
$ where $a(t)$ is the scale factor and $H=da/dt$ is the Hubble parameter. We
also assume the inflaton $\phi $ that only depends on time $t$. This field
is similar to an axion, a light pseudoscalar particle with a shift symmetry $%
\phi \rightarrow \phi +\phi _{0}$. We take a simple quadratic potential for
the inflaton, $V\left( \phi \right) =\frac{1}{2}m^{2}\phi ^{2}$, where $%
m=1.9\times 10^{13}GeV\left( \propto \Lambda ^{2}\right) $ \cite{JCAP5}. The
background dynamics of $\phi \left( t\right) $ and $A(t,x)$ \cite{JCAP2} are
given by
\begin{equation}
\ddot{\phi}+3H\dot{\phi}+\frac{dV\left( \phi \right) }{d\phi }-\frac{\nabla
^{2}\phi }{a^{2}\left( t\right) }=\frac{\sigma }{\Lambda }\left\langle
\mathbf{E\cdot B}\right\rangle ,  \label{m2}
\end{equation}%
\begin{equation}
3M_{p}^{2}H^{2}-\frac{1}{2}\dot{\phi}^{2}-V\left( \phi \right) -\frac{\left(
\nabla \phi \right) ^{2}}{2a^{2}\left( t\right) }=\frac{1}{2}\left\langle
\mathbf{E}^{2}+\mathbf{B}^{2}+\frac{m_{A}^{2}}{a^{2}}\mathbf{A}%
^{2}\right\rangle ,  \label{m3}
\end{equation}%
where $\mathbf{E}=-\frac{\mathbf{A}\prime }{a^{2}}=-\frac{1}{a^{2}}\frac{%
\partial \mathbf{A}}{\partial \tau }$ and $\mathbf{B}=\frac{\mathbf{\nabla }%
\times \mathbf{A}}{a^{2}}$ are the electric and magnetic fields
corresponding to the gauge field respectively. Here $M_{p}^{2}\approx
2.4\times 10^{18}GeV$ is the reduced Planck mass. We now consider that the
evolution of the scale factor is written as a power law $a=a_{0}t^{q}$. That
implies the Hubble parameter can rewrite as $H=\frac{1}{a\tau }$, where $%
\tau $ is the conformal time with $dt=a\left( \tau \right) d\tau $. The
inflaton perturbations follow the equation of motion%
\begin{equation}
\delta \ddot{\phi}+3\beta H\delta \dot{\phi}-\left( \frac{\nabla ^{2}}{a^{2}}%
-\frac{d^{2}V\left( \phi \right) }{d\phi ^{2}}\right) \delta \phi =\frac{%
\sigma }{\Lambda }\left( \mathbf{E\cdot B-}\left\langle \mathbf{E\cdot B}%
\right\rangle \right) .  \label{jj2}
\end{equation}%
We define the curvature perturbation on hypersurfaces characterized by
uniform density as $\zeta \left( \tau ,\mathbf{x}\right) =-\frac{H}{\dot{\phi%
}}\delta \phi \left( \tau ,\mathbf{x}\right) $. Without the presence of
gauge fields, the inflaton perturbation's classical vacuum solution can be
formulated as follows: $\delta \phi \left( \tau ,\mathbf{x}\right) =\frac{H}{%
\sqrt{2k^{3}}}\left( 1+ik\tau \right) e^{-ik\tau }$, and we take
\begin{equation}
\zeta \left( \tau ,\mathbf{x}\right) =-\frac{H^{2}}{\dot{\phi}\sqrt{2k^{3}}}%
\left( 1+ik\tau \right) e^{-ik\tau }.  \label{jj3}
\end{equation}%
The decoupling of modes can be achieved by breaking down the quantum field $%
\mathbf{A}\left( \tau ,\mathbf{x}\right) $ into components:%
\begin{eqnarray}
\mathbf{A}\left( \tau ,\mathbf{x}\right)  &=&\sum_{\lambda =\pm ,0}\int
\frac{d^{3}k}{\left( 2\pi \right) ^{3}}\mathbf{\epsilon }_{\lambda }\left(
\mathbf{k}\right) e^{i\mathbf{k\cdot x}}(a_{\lambda }\left( \mathbf{k}%
\right) A_{\lambda }\left( \tau ,k\right)   \label{m4} \\
&&+a_{-\lambda }^{\dagger }\left( \mathbf{k}\right) A_{\lambda }^{\ast
}\left( \tau ,k\right) .)  \notag
\end{eqnarray}%
The polarization vector, $\mathbf{\epsilon }_{\lambda }\left( \mathbf{k}%
\right) $, and the annihilation operator, $a_{\lambda }\left( \mathbf{k}%
\right) $, adhere to the standard commutation and orthonormality relations.
The commutation relation of the creation ($a_{-\lambda }^{\dagger }$) and
annihilation ($a_{\lambda }\left( \mathbf{k}\right) $) operators can be
expressed as $\left[ a_{\lambda }\left( \mathbf{k}\right) ,a_{-\lambda
^{\prime }}^{\dagger }\left( \mathbf{k}^{\prime }\right) \right] =\left(
2\pi \right) ^{3}\delta _{\lambda \lambda ^{\prime }}\delta ^{\left(
3\right) }\left( \mathbf{k}-\mathbf{k}^{\prime }\right) .$ The generation of
the dominant vector field is controlled by the field equations governing the
transverse Fourier modes%
\begin{equation}
\frac{\partial ^{2}A_{\mathbf{\pm }}\left( \tau ,k\right) }{\partial \tau
^{2}}+\left( k^{2}+a^{2}\left( \mathbf{\tau }\right) m_{A}^{2}\pm \frac{%
\sigma k\dot{\phi}}{\Lambda H\tau }\right) A_{\pm }\left( \tau ,k\right) =0.
\label{m5}
\end{equation}%
where $\dot{\phi}=\frac{\partial \phi }{\partial \tau }$, $A_{\pm }\left(
\tau ,k\right) $ represents the mode function, with the longitudinal mode
denoted as $\lambda =0$ and the two transverse modes as $\lambda =\pm $,
respectively. The expression for the two helicity modes function of the
vector field can be formulated using the $W_{\mp i\xi ,i\omega }$ as%
\begin{equation}
A_{\mathbf{\pm }}\left( \tau ,k\right) =\frac{1}{\sqrt{2k}}e^{\pm \frac{\pi
\xi }{2}}W_{\mp i\xi ,i\omega }\left( 2ik\tau \right) .  \label{m6}
\end{equation}%
where $W_{\mp i\xi ,i\omega }$ is the Whittaker W function. Here we have
introduced a convenient notation%
\begin{eqnarray}
\xi  &=&\frac{\sigma \dot{\phi}}{2\Lambda H},  \label{m8} \\
\omega ^{2} &=&\frac{m_{A}^{2}}{H^{2}}-\frac{1}{4}.  \label{m9}
\end{eqnarray}%
The Taylor expansion of the Whittaker function around a point $2ik\tau _{0}$
can be expressed as follows:%
\begin{eqnarray}
W_{\mp i\xi ,i\omega }\left( 2ik\tau \right)  &=&e^{-ik\tau }z^{i\omega +%
\frac{1}{2}}\sum_{n=0}^{\infty }\frac{\left( -1\right) ^{n}}{n!}  \label{m10}
\\
&&\times \frac{\Gamma \left( i\omega \mp i\xi +\frac{1}{2}-n\right) }{\Gamma
\left( i\omega \mp i\xi +\frac{1}{2}\right) }\left( 2ik\tau -2ik\tau
_{0}\right) ^{n}.  \notag
\end{eqnarray}%
In this case, we can deduce the expression for $A_{\mathbf{\pm }}$ as%
\begin{eqnarray}
A_{\mathbf{\pm }} &=&\frac{z^{i\omega +\frac{1}{2}}}{\sqrt{2k}}e^{\pm \frac{%
\pi \xi }{2}-ik\tau }\sum_{n=0}^{\infty }\frac{\left( -1\right) ^{n}\left(
2ik\right) ^{n}}{n!}  \label{m11} \\
&&\times \frac{\Gamma \left( i\omega \mp i\xi +\frac{1}{2}-n\right) }{\Gamma
\left( i\omega \mp i\xi +\frac{1}{2}\right) }\left( \tau -\tau _{0}\right)
^{n}.  \notag
\end{eqnarray}%
We can use this approximation $A_{\mathbf{\pm }}\sim \frac{z^{i\omega +\frac{%
1}{2}}}{\sqrt{2k}}e^{\pm \frac{\pi \xi }{2}-ik\tau }\left[ 1-2ik\left( \tau
-\tau _{0}\right) \frac{\Gamma \left( i\left( \omega \mp \xi \right) -\frac{1%
}{2}\right) }{\Gamma \left( i\left( \omega \mp \xi \right) +\frac{1}{2}%
\right) }\right] $\ to plot Figs. (\ref{F1})-(\ref{F2})-(\ref{F3}). For $%
\tau =\tau _{0}$, the gauge field can be expressed as%
\begin{equation}
A_{\mathbf{\pm }}\left( \tau _{0}\right) \sim \frac{1}{\sqrt{2k}}z^{i\omega +%
\frac{1}{2}}e^{\pm \frac{\pi \xi }{2}-ik\tau _{0}}.  \label{m13}
\end{equation}%
We can relate the gauge field to the Hubble parameter with this ratio $\frac{%
A_{\mathbf{+}}\left( \tau _{0},k\right) }{A_{\mathbf{-}}\left( \tau
_{0},k\right) }\sim e^{\pi \xi }$. In the Figs. (\ref{F1})-(\ref{F2})-(\ref%
{F3}) we will only study the evolution of $A_{\mathbf{+}}$ in term of $%
\left( \omega ,\xi \right) $.
We generate a grid of $\omega $ and $\xi $ values to compute $A_{\mathbf{+}}$
over an interval in Fig. \ref{F1}. As shown in Figs. (\ref{F1})-(\ref{F2}),
the parameter $\tau _{0}$ is crucial for describing how $A_{\mathbf{+}}$
changes. Moreover, the diagram in Fig. (\ref{F3}) not only shows the case
for $\tau _{0}=3$, but also illustrates how $A_{\mathbf{+}}$ evolves for all
other values of $\tau _{0}\geq 3$. The goal of this study is to examine how
the mass of the gauge field $m_{A}\sim \omega $ affects the Hubble parameter
$H\sim \frac{1}{\xi }$. Perhaps the discrepancy in the measurements of the
universe's expansion rate, known as the Hubble tension \cite{HT1,HT2,HT3},
is related to how the Hubble parameter $H$ depends on the mass of the gauge
bosons in the early universe. The value of $\tau _{0}=3$ signifies the
critical point at which the mass of the gauge boson begins to affect the
Hubble parameter. This effect is primarily due to the space-time curvature
induced by these massive bosons.

\textit{Hopf fibration}. In the study \cite{AX6}, they showed the existence
of gravitational type gravitational hopfions. Likewise, we will consider
that the additional term $M_{\mp i\xi ,-i\omega }$ which is added to $A_{%
\mathbf{\pm }}$ describes the hopfions in the early universe. In this
section, our objective is to demonstrate the presence of hopfions in the
early universe. The average energy density of the gauge field $A_{\mu }$ can
be written as $\rho _{A}=\frac{1}{2}\left\langle \mathbf{E}^{2}+\mathbf{B}%
^{2}+\frac{m_{A}^{2}}{a^{2}}\mathbf{A}^{2}\right\rangle .$ We express the
term $\mathbf{E\cdot B}$ in the form given by \cite{JCAP5}%
\begin{equation}
\left\langle \mathbf{E\cdot B}\right\rangle =-\frac{1}{a^{4}}\int \frac{%
d^{3}k}{\left( 2\pi \right) ^{3}}\frac{k}{2}\frac{d}{d\tau }\left(
\left\vert A_{\mathbf{+}}\right\vert ^{2}-\left\vert A_{\mathbf{-}%
}\right\vert ^{2}\right) .  \label{p1}
\end{equation}%
Next, we introduce the conjugate term:%
\begin{equation}
\left\langle \mathbf{E\cdot B}\right\rangle ^{\ast }\equiv -\frac{1}{a^{4}}%
\int \frac{d^{3}k}{\left( 2\pi \right) ^{3}}\frac{k}{2}\frac{d}{d\tau }%
\left( 2A_{\mathbf{+}}A_{\mathbf{-}}^{\ast }\right) .  \label{p2}
\end{equation}%
The spherically equivariant map to compactifiy $%
\mathbb{R}
^{3}\rightarrow S_{A_{\mathbf{+}},A_{\mathbf{-}}}^{3}\subset
\mathbb{C}
^{2}$ is
\begin{equation}
S_{A_{\mathbf{+}},A_{\mathbf{-}}}^{3}\cong \left\{ \left( A_{\mathbf{+}}A_{%
\mathbf{-}}\right) \in
\mathbb{C}
^{2},\left\vert A_{\mathbf{+}}\right\vert ^{2}+\left\vert A_{\mathbf{-}%
}\right\vert ^{2}=1\right\} ,
\end{equation}%
where $S_{A_{\mathbf{+}},A_{\mathbf{-}}}^{3}\simeq SU(2)$. The Hopf
fibration $\mathbf{P}_{h}$ is then defined as \cite{HF0,HF1,HF2}
\begin{equation}
\mathbf{P}_{h}\left( A_{\mathbf{+}},A_{\mathbf{-}}\right) =\left( 2A_{%
\mathbf{+}}A_{\mathbf{-}}^{\ast },\left\vert A_{\mathbf{+}}\right\vert
^{2}-\left\vert A_{\mathbf{-}}\right\vert ^{2}\right) ,  \label{hp}
\end{equation}%
which satisfies $\frac{d\mathbf{P}_{h}\left( A_{\mathbf{+}},A_{\mathbf{-}%
}\right) }{d\tau }=-3\left( 2\pi \right) ^{3}\left( \frac{2a}{k}\right)
^{4}\left( \left\langle \mathbf{E\cdot B}\right\rangle ^{\ast },\left\langle
\mathbf{E\cdot B}\right\rangle \right) .$ The Hopf index $\mathcal{H}$
emerges as the most appropriate topological invariant for characterizing
these phenomena%
\begin{equation}
\mathcal{H}=\frac{1}{\left( 8\pi \right) ^{2}}\int d^{3}x\mathbf{A\cdot B.}
\label{p5}
\end{equation}%
The spacetime Hopf index $\mathcal{H}$ emerges as the most appropriate
topological invariant for characterizing these phenomena $\frac{d\mathcal{H}%
}{d\tau }=-\frac{\pi a^{2}}{\left( 2\pi \right) ^{3}}\int d^{3}x\mathbf{%
E\cdot B-}\frac{\pi a}{\left( 2\pi \right) ^{3}}\int d^{3}x\mathbf{A\cdot }%
\nabla \times \mathbf{E,}$ where $\frac{\partial \mathbf{B}}{\partial t}=%
\frac{\partial \mathbf{B}}{a\partial \tau }=-\nabla \times \mathbf{E}$. This
perturbation can be expressed as
\begin{equation}
\frac{d\mathcal{H}}{d\tau }\mathbf{\approx }-\frac{\pi a^{2}}{\left( 2\pi
\right) ^{3}}\int d^{3}x\left\langle \mathbf{E\cdot B}\right\rangle .
\label{p7}
\end{equation}%
Substituting Eq. (\ref{p1}) into Eq. (\ref{p7}), we ascertain that
\begin{equation*}
\frac{d\mathcal{H}}{d\tau }\mathbf{=}\frac{\pi }{2\left( 2\pi \right)
^{3}a^{2}}\int d^{3}x\int \frac{d^{3}k}{\left( 2\pi \right) ^{3}}k\frac{d}{%
d\tau }\left( \left\vert A_{\mathbf{+}}\right\vert ^{2}-\left\vert A_{%
\mathbf{-}}\right\vert ^{2}\right) e^{ikx}e^{-ikx},
\end{equation*}%
or equivalently%
\begin{equation}
\frac{d\mathcal{H}}{d\tau }\mathbf{=}\frac{\pi k}{2\left( 2\pi \right)
^{3}a^{2}}\frac{d}{d\tau }\left( \left\vert A_{\mathbf{+}}\right\vert
^{2}-\left\vert A_{\mathbf{-}}\right\vert ^{2}\right) .  \label{p9}
\end{equation}%
Subsequently we assume that $a^{2}=\frac{\pi k}{2\left( 2\pi \right) ^{3}}$
and we find%
\begin{equation}
\mathcal{H}\mathbf{=}\left\vert A_{\mathbf{+}}\right\vert ^{2}-\left\vert A_{%
\mathbf{-}}\right\vert ^{2}.  \label{p10}
\end{equation}%
The terms $\left\vert A_{\mathbf{+}}\right\vert $ and $\left\vert A_{\mathbf{%
-}}\right\vert ^{2}$ can be linked to energy or load density associated with
these fields. From this relationship we notice for $\mathcal{H}=0$, the
magnitudes of $\left\vert A_{\mathbf{+}}\right\vert $ and $\left\vert A_{%
\mathbf{-}}\right\vert $ are equal. On the other hand we can calculate the
conjugate of $\mathcal{H}$ as follows%
\begin{equation}
\frac{d\mathcal{H}^{\ast }}{d\tau }=-\frac{\pi a^{2}}{\left( 2\pi \right)
^{3}}\int d^{3}x\left\langle \mathbf{E\cdot B}\right\rangle ^{\ast }.
\label{p11}
\end{equation}%
Replacing the expression from Eq. (\ref{p2}) into Eq. (\ref{p11}), we
establish that
\begin{equation*}
\frac{d\mathcal{H}^{\ast }}{d\tau }=\frac{\pi k}{2a^{2}\left( 2\pi \right)
^{3}}\frac{d}{d\tau }\left( 2A_{\mathbf{+}}A_{\mathbf{-}}^{\ast }\right) .
\end{equation*}%
Setting $a^{2}=\frac{\pi k}{2\left( 2\pi \right) ^{3}}$, we derive that
\begin{equation}
\mathcal{H}^{\ast }=2A_{\mathbf{+}}A_{\mathbf{-}}^{\ast }.  \label{p12}
\end{equation}%
\begin{figure}
\centering\includegraphics[width=10cm]{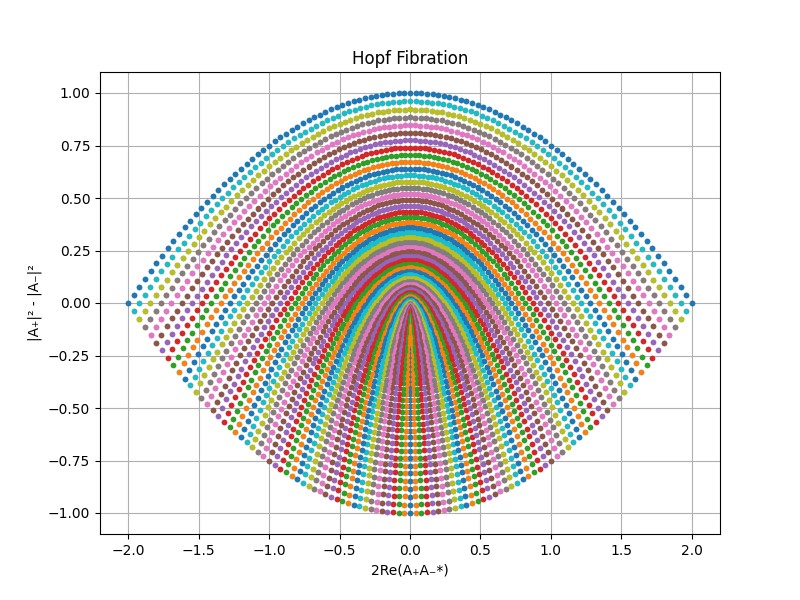}
\caption{This figure represents the Hopf fibration $\mathbf{P}_{h}$ about
the gauge field modes $A_{\mathbf{+}}$ and $A_{\mathbf{-}}$. Each point on
the plot signifies the outcome of the Hopf fibration for a specific pair of $%
A_{\mathbf{+}}$ and $A_{\mathbf{-}}$ values. }
\label{F4}
\end{figure}
Fig. (\ref{F4}) illustrates the outcome of our research, demonstrating that
the massive gauge field can produce a Hopf fibration, representing regions
in space-time where Hopfions are applicable. These Hopfions encompass points
that form lines, adhering to the space-time curvature induced by the mass $%
m_{A}$. From Eqs. (\ref{hp})-(\ref{p10})-(\ref{p12}) we get%
\begin{equation}
\mathbf{P}_{h}\left( A_{\mathbf{+}},A_{\mathbf{-}}\right) =\left( \mathcal{H}%
^{\ast },\mathcal{H}\right) .  \label{p13}
\end{equation}%
We have shown that the Hopf fibration is articulated through the presence of
two Hopf index components Fig. (\ref{F4}). This revelation enables the
characterization of the Hopfion state in the early universe. Following the
inflationary epoch characterized by the inflation field, the radiation era
begins, orchestrated by the electromagnetic field. However, the emergence of
topological defects in spacetime, specifically hopfions, occurs during this
phase. Within these regions, novel fluctuations manifest, potentially
serving as the source of the pulsar timing arrays signal \cite%
{NANO0,NANO00,NANO1,NANO2}. On the other hand, the $\mathbf{P}_{h}\left( A_{%
\mathbf{+}},A_{\mathbf{-}}\right) $ resides on the unit 2-sphere in $\mathbb{%
C} \times\mathbb{R}$. \textit{Conclusion}. This study introduces a new
method for examining the dynamics of inflation fields in conjunction with a
massive gauge boson. We developed a new model that characterizes the
curvature of space-time, influenced by a massive gauge field in the early
universe. This curvature can lead to numerous observations, such as the
Hubble tension issue and the isotropic stochastic gravitational-wave
background. To address this, we are introducing the concept of gauge field
Hopfions. The solution for the gauge field is represented by the Whittaker
function. This framework allows for the direct determination of the
relationships between the mode functions of the gauge field $A_{\mathbf{\pm }%
}$, the Hubble parameter $H$, and the mass of the gauge field $m_{A}$.
Additionally, our results show a significant correlation between the Hubble
parameter and the mass $m_{A}$, as shown in the curves Figs. (\ref{F1})-(\ref%
{F2})-(\ref{F3}), which illustrate that the parameter $\tau _{0}$ is crucial
for determining the variation of mode $A_{\mathbf{+}}$. This correlation
could potentially offer a comprehensive explanation for the Hubble tension.
Moreover, Fig. (\ref{F3}) presents the case for $\tau _{0}=3$ and also the
evolution of mode $A_{\mathbf{+}}$ for all other values of $\tau \geq 3$.
The goal of this research is to investigate how the Hubble parameter is
affected by the mass of the gauge field. Furthermore, we have shown that the
mode functions of the gauge field can be expressed within the context of a
component of the Hopf fibration. After our calculations, we confirmed the
accuracy of this proposition by demonstrating that the Hopf fibration of the
gauge field is articulated in terms of the Hopf index $\mathcal{H}$ and
their combination. We have demonstrated that the Hopf fibration is expressed
via the existence of two components of the Hopf index as shown in Fig. (\ref%
{F4}). This discovery allows for the description of the Hopfion state in the
early universe. After the inflationary period defined by the inflation
field, the era of radiation begins, governed by the electromagnetic field.
However, during this phase, topological defects in space-time, particularly
hopfions, emerge. Within these areas, new fluctuations appear, which could
potentially be the origin of the pulsar timing arrays.

\end{document}